\documentclass{svproc}
\usepackage{url}

\usepackage{amsmath}
\usepackage{amssymb}
\usepackage{calc}
\usepackage{siunitx}
\usepackage{pgfplots}
\usepgfplotslibrary{colormaps}
\usetikzlibrary{arrows.meta, positioning, calc, fit}
\usetikzlibrary[pgfplots.groupplots]
\pgfplotsset{compat=newest}
\pgfplotsset{
	colormap={pv-rainbow}{
		rgb255=(5, 97, 255) rgb255=(5, 112, 244) rgb255=(5, 127, 234)
		rgb255=(5, 140, 222) rgb255=(5, 152, 209) rgb255=(5, 164, 194)
		rgb255=(5, 176, 178) rgb255=(5, 188, 161) rgb255=(5, 201, 142)
		rgb255=(5, 214, 123) rgb255=(6, 225, 100) rgb255=(4, 237, 76)
		rgb255=(84, 244, 36) rgb255=(148, 248, 18) rgb255=(193, 251, 9)
		rgb255=(234, 254, 5) rgb255=(255, 247, 15) rgb255=(255, 232, 38)
		rgb255=(255, 216, 55) rgb255=(255, 199, 55) rgb255=(255, 183, 55)
		rgb255=(255, 166, 55) rgb255=(255, 150, 55) rgb255=(255, 131, 55)
		rgb255=(255, 111, 55) rgb255=(254, 87, 54) rgb255=(252, 60, 50)
		rgb255=(247, 33, 60) rgb255=(231, 21, 74) rgb255=(214, 10, 86)
		rgb255=(194, 12, 96) rgb255=(174, 13, 106)
	},
	colormap name=pv-rainbow
}
\pgfplotsset{
	colormap={pv-bluewhitered}{rgb255=(59,76,193) rgb255=(221,221,221) rgb255=(180,4,39)
	},
	colormap name=pv-bluewhitered
}

\usepackage{eso-pic}
\AddToShipoutPictureFG*{
	\AtPageUpperLeft{
		\raisebox{-5\baselineskip}{
			\makebox[\paperwidth]{
				\begin{minipage}{21cm}
					\centering
					\color{gray}
					This is a preprint of the following chapter:\\
					Florian J\"ackel,\\
					Sensitivity Analysis of Discrepancy Terms Introduced in Turbulence Models Using Field Inversion,\\
					published in New Results in Numerical and Experimental Fluid Mechanics XIII,\\
					edited by Prof. Dr. Dr. Andreas Dillmann, Dr. Gerd Heller, Prof. Dr. Ewald Kr\"amer, Prof. Dr. Claus Wagner, 2021, Springer,\\
					reproduced with permission of Springer International Publishing.\\
					The final authenticated version is available online at: \url{http://dx.doi.org/10.1007/978-3-030-79561-0_59}
				\end{minipage}}}
			}
}

\graphicspath{{figures/}}

\begin{document}
\mainmatter              
\title{Sensitivity Analysis of Discrepancy Terms introduced in Turbulence Models using Field Inversion}
\titlerunning{Sensitivity Analysis Field Inversion}  
%
\author{Florian J\"ackel\inst{1}}
\authorrunning{Florian J\"ackel} 
%
\tocauthor{Florian J\"ackel}
\institute{Institute of Aerodynamics and Flow Technology, Center for Computer Applications in AeroSpace Science and Engineering, Bunsenstra{\ss}e 10, 37073 G\"ottingen, Germany,\\\email{florian.jaeckel@dlr.de}}

\maketitle              

\begin{abstract}
	RANS simulations with the Spalart-Allmaras turbulence \linebreak[4]model are improved for cases with flow separation using the Field Inversion and Machine Learning approach.
	A compensatory discrepancy term is introduced into the turbulence model and optimized using high-fidelity reference data from experiments.
	Influences on the optimization results with respect to regularization, grid resolution and areas in which the optimization is active are investigated.
	Finally, a neural network is trained and used to augment simulations on a test case.
\end{abstract}
\section{Introduction}
Turbulent flows often play a key role in engineering and in science, hence it is important to be able to predict these flows accurately by numerical simulation.
Methods which can resolve turbulent scales like LES and DNS are often still too expensive to be used regularly. Hence, Reynolds-Averaged Navier Stokes (RANS) simulations, which model the influence of turbulence on the mean flow, are the standard method for daily use.
One difficulty however is that RANS simulations often fail to predict flow separation accurately e.g. in high-lift flows.

Over the last decades, many approaches aiming to improve turbulence modeling have been tried with varying degrees of success, such as extending existing models, for example with rotation corrections, or using more complex turbulence models, e.g. two equation models or Reynolds-stress models.
One approach that has gained increasing interest in the recent years is the Field Inversion and Machine Learning (FIML) approach \cite{fiml}.
For this approach, it is argued that RANS models have always been largely empirical, and hence, with the increasing maturity of machine learning techniques, they can further be improved using machine learning models trained on datasets derived from high-fidelity simulations like DNS and LES or from experiments.

In this paper, an implementation of the FIML approach based on the negative Spalart-Allmaras turbulence model \cite{SAneg} in the DLR TAU-Code is described.
In particular, findings on the influence of the regularization, grid resolutions and regions of the computational domain where the field inversion is active on the field inversion results are presented. 
Finally, the turbulence model is augmented with a machine learning model and improved results computed with the augmented model are shown.%
\section{Methodology}
\label{chap:methodolgy}
\subsection{Field Inversion and Machine Learning}
The basic idea of the FIML approach is to introduce a discrepancy term $\beta$ into the turbulence model, here the negative Spalart-Allmaras model:
\begin{align}
	\frac{D \tilde{\nu}}{D t} 
		&= \beta(\mathbf{U}, \tilde{\nu}) P(\mathbf{U}, \tilde{\nu}) - D(\mathbf{U}, \tilde{\nu}) + T(\mathbf{U}, \tilde{\nu})
	\label{eq:extended_turb_model}
\end{align}
Here, $P, D$ and $T$ denote the turbulent production, destruction and transport terms respectively and $\mathbf{U}$ and $\tilde{\nu}$ denote the flow state and the Spalart-Allmaras transport variable.
The correction term $\beta$ is an unknown, spatially varying variable which is assumed to be a function of $\mathbf{U}$ and $\tilde{\nu}$.
Since $\beta$ is unknown, it must be modeled as well.
The modeling is done in two separate steps: During a \textit{field inversion} step training data is generated which is then used in a \textit{machine learning} step to train a model relating $\beta$ and $(\mathbf{U},\tilde{\nu})$.

\paragraph{Field Inversion.}
The training data needed for machine learning consists of pairs of $\beta$ and the corresponding flow state $(\mathbf{U}, \tilde{\nu})$, one pair per control volume.
These pairs are obtained from high-fidelity reference data $d_\text{ref}$ by solving the inverse problem
\begin{equation}
	d_\text{RANS}(\beta) = d_\text{ref},
	\label{eq:inverse_problem}
\end{equation}
where $d_\text{RANS}$ is the RANS solution depending on $\beta$, for $\beta$. 
Solving Eq. (\ref{eq:inverse_problem}) is an optimization problem, for which a cost function, here
\begin{equation}
	\mathcal{I} = 
		\underbrace{\frac{1}{2} \sum_{i}^{N_i} \sum_{j}^{N_{j,i}} \left[ d_{i, \text{ref}}^j - d_{i, \text{RANS}}^j(\beta, \mathbf{U}, \tilde{\nu}) \right]^2}_{\mathcal{I}_1}
		+ \lambda \underbrace{\frac{1}{2} \sum_{k}^{N_k} \left( \beta_k - 1 \right)^2}_{\mathcal{I}_2}
	\label{eq:cost_fn}
\end{equation}
needs to be minimized.
The first term of the cost function, $\mathcal{I}_1$, measures the deviation between $d_\text{ref}$ and $d_\text{RANS}$ and is calculated as the squared difference of $d_\text{ref}$ and $d_\text{RANS}$, summed over all $N_{j,i}$ cells where reference data is given and over all $N_i$ reference quantities.
Reference quantities can be field variables, such as velocities $u$, surface variables, such as the skin friction $c_f$, or integral values such as the lift coefficient $c_l$.

The second term, $\mathcal{I}_2$, measures and penalizes the magnitude of the turbulence model modification, acting as a Tikhonov regularization.
Regularization is needed as the optimization problem typically is ill-posed due to noise in the reference data and the degrees of freedom, i.e. number of cells where $\beta$ must be optimized, being much higher than the number of cells where reference data is given.
The regularization parameter $\lambda$ is selected according to the \textit{L-Curve} criterion \cite{hansen}.

Because of the non-linearities in the RANS equations which are contained in $\mathcal{I}$, minimization of $\mathcal{I}$ with respect to $\beta$ must be done iteratively.
Hence, a gradient descent method is employed for which the gradient can be computed efficiently using the adjoint method.

\paragraph{Machine Learning.}
The field inversion step returns discrete values of $\beta$ at the grid nodes, which can't be transferred directly to simulations on different geometries or different flow conditions.
Hence, a relation between $\beta$ and $(\mathbf{U}, \tilde{\nu})$ must be found first.
Therefore, dimensionless flow features $\eta_i(\mathbf{U}, \tilde{\nu})$ are derived and then machine learning is used to find a model
\begin{equation}
	f_\beta: \eta_0(\mathbf{U}, \tilde{\nu}), \dots, \eta_n \mapsto \beta
	\label{eq:ml-beta}
\end{equation} 
approximating $\beta$.
In particular, neural networks are used for this regression task.
Due to the limited space here, the interested reader is referred to dedicated publications such as \cite{deeplearning} for a detailed description of the principles of neural networks.

\subsection{Implementation}

The Field Inversion and Machine Learning approach has been implemented using the DLR TAU code \cite{TAU}, a highly optimized, parallel, state of the art CFD solver for unstructured grids, using and extending its adjoint capabilities.
TensorFlow \cite{tensorflow} is used for training and evaluating the machine learning model.
The negative Spalart-Allmaras (SA-neg) \cite{SAneg} model is used as turbulence model due to its focus on aeronautical boundary layer flow.
\section{Sensitivity Analysis}
\label{chap:numerical_results}

For the following sections, field inversions are conducted on the S809 airfoil.
This airfoil was developed at the National Renewable Energy Laboratory (NREL) for wind turbines, and pressure distributions from wind tunnel measurements are available for a wide range of angles of attack and for multiple Reynolds numbers \cite{s809}.
At an angle of attack of $\alpha \approx 6^\circ$, flow separation starts to develop at the trailing edge and the predictions of the RANS simulations become unreliable, hence making this case suitable for the FIML approach, where it has been used successfully before \cite{fiml}.
The field inversions in this section are based on a case with an angle of attack of $\alpha=12.2^\circ$, a Mach number of Ma = 0.1 and a Reynolds number of Re  = $2 \times 10^6$ with the experimentally measured surface distribution of $c_p$ from \cite{s809} as reference.

\subsection{Regularization}

\begin{figure}
	\begin{minipage}[t]{.45\textwidth}
		\def\figwidth  {\textwidth}
		\def\figheight {1.2\textwidth}
\begin{tikzpicture}

\begin{axis}[
axis equal,
height=\figheight,
legend pos=south west,
log basis x={10},
log basis y={10},
minor xtick={0.02, 0.03, 0.04, 0.05, 0.06, 0.07, 0.08, 0.09, 0.2, 0.3, 0.4, 0.5, 0.6, 0.7, 0.8, 0.9},
minor ytick={10, 20, 30, 40, 50, 60, 70, 80, 90, 200, 300, 400, 500, 600, 700, 800, 900, 2000, 3000},
tick pos=left,
width=\figwidth,
x grid style={white!69.0196078431373!black},
xlabel={\(\displaystyle \mathcal{I}_1\)/\(\displaystyle \mathcal{I}_{1,0}\)},
xmajorgrids,
xmin=0.0332138503407102, xmax=0.219207893138161,
xmode=log,
xtick style={color=black},
xtick={0.01, 0.1, 1.},
y grid style={white!69.0196078431373!black},
y label style={at={(axis description cs:-0.1,.5)},rotate=-90},
ylabel={\(\displaystyle \mathcal{I}_2\)},
ymajorgrids,
ymin=137.700931992188, ymax=1285.68538514185,
ymode=log,
ytick style={color=black},
ytick={100, 1000, 1000}
]
\addplot [semithick, black, mark=x, mark size=3, mark options={solid,fill opacity=0}]
table {%
0.201189127770607 152.41818331257
};
\addplot [semithick, black, mark=x, mark size=3, mark options={solid,fill opacity=0}]
table {%
0.0741375103306659 432.31190105015
};
\addplot [semithick, black, mark=x, mark size=3, mark options={solid}]
table {%
0.0515849168638916 536.228765584584
};
\addplot [semithick, black, mark=x, mark size=3, mark options={solid,fill opacity=0}]
table {%
0.0411600271644161 645.520612606546
};
\addplot [semithick, black, mark=x, mark size=3, mark options={solid,fill opacity=0}]
table {%
0.0390371535674743 702.506849162487
};
\addplot [semithick, black, mark=x, mark size=3, mark options={solid,fill opacity=0}]
table {%
0.037398437202144 798.695359487336
};
\addplot [semithick, black, mark=x, mark size=3, mark options={solid,fill opacity=0}]
table {%
0.0361885268695765 1161.54170017697
};
\addplot [semithick, black, dotted]
table {%
0.201189127770607 152.41818331257
0.0741375103306659 432.31190105015
0.0515849168638916 536.228765584584
0.0411600271644161 645.520612606546
0.0390371535674743 702.506849162487
0.037398437202144 798.695359487336
0.0361885268695765 1161.54170017697
};
\draw (axis cs:0.201189127770607,152.41818331257) node[
  anchor= east,
  text=black,
  rotate=0.0
]{\,$\num{2e-3}\,$};
\draw (axis cs:0.0741375103306659,432.31190105015) node[
  anchor= west,
  text=black,
  rotate=0.0
]{\,$\num{1e-3}\,$};
\draw (axis cs:0.0515849168638916,536.228765584584) node[
  anchor= west,
  text=black,
  rotate=0.0
]{\,$\num{5e-4}\,$};
\draw[-latex,very thick,draw=black] (axis cs:0.0515849168638916,236.228765584584) -- (axis cs:0.0515849168638916,486.228765584584);
\draw (axis cs:0.0411600271644161,645.520612606546) node[
  anchor= west,
  text=black,
  rotate=0.0
]{\,$\num{2e-4}\,$};
\draw (axis cs:0.0390371535674743,702.506849162487) node[
  anchor=south west,
  text=black,
  rotate=0.0
]{\,$\num{1e-4}\,$};
\draw (axis cs:0.037398437202144,798.695359487336) node[
  anchor=south west,
  text=black,
  rotate=0.0
]{\,$\num{1e-5}\,$};
\draw (axis cs:0.0361885268695765,1161.54170017697) node[
  anchor=south west,
  text=black,
  rotate=0.0
]{\,$\lambda = \num{1e-6}\,$};
\end{axis}

\end{tikzpicture}
	\end{minipage}%
	\hfill
	\begin{minipage}[t]{.55\textwidth}
		\def\figwidth {\textwidth}
		\def\figheight {0.9818\textwidth}
		\input{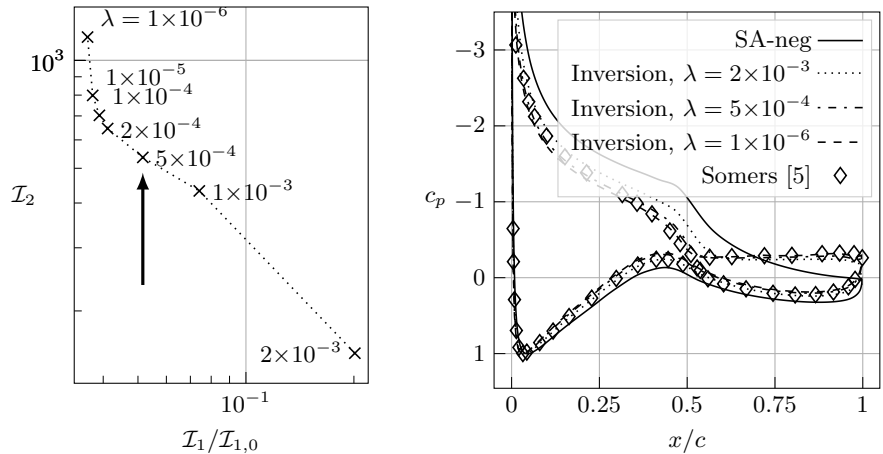}
	\end{minipage}
	\caption{Left: The L-Curve for the field inversion results for the S809. The optimal $\lambda$ is marked by the arrow. Right: $c_p$-distributions from field inversions for different $\lambda$.}
	\label{fig:s809_reg}
\end{figure}

As described in Chapter \ref{chap:methodolgy}, the present optimization problem is ill-posed and therefore requires regularization.
For an optimal choice of the parameter $\lambda$, the \textit{L-Curve} criterion \cite{hansen} is considered.
Hence, field inversions are conducted for a range of values for $\lambda$ and the resulting cost function terms $\mathcal{I}_1$, $\mathcal{I}_2$ are plotted against each other in the log-log plot in Figure \ref{fig:s809_reg}, on the left.
The optimal $\lambda_\text{opt}$ is found at the curve's inflection point, indicating the point beyond which $\mathcal{I}_1$ does not decrease significantly anymore but the magnitude of the model modification represented by $\mathcal{I}_2$ increases fast.
For the current case, the optimal value for the regularization parameter is $\lambda = \num{5e-4}$ as indicated by the arrow in Figure \ref{fig:s809_reg}.

The right side of Figure \ref{fig:s809_reg} shows the $c_p$ distributions of the baseline turbulence model, the field inversion results for $\lambda = \lambda_\text{opt}$, a $\lambda < \lambda_\text{opt}$ and a $\lambda > \lambda_\text{opt}$ and the reference data.
The reference data as well as the field inversion results exhibit a pressure plateau beginning at $x/c \approx 0.5$, indicating flow separation at the trailing edge which is not predicted by the baseline model.
For $\lambda = \lambda_\text{opt} = \num{5e-4}$, the $c_p$-distribution (solid line) is close to the reference data and the value of cost function part $\mathcal{I}_1$ has decreased to $5.16\%$ of its initial value.
For stronger regularization, here e.g. $\lambda = \num{2e-3}$, the $c_p$-distribution (dash-dotted line) deviates further from the reference data with a decrease of cost function $\mathcal{I} _1$ to $20.1\%$ of its initial value.
A pressure plateau is still developed, but is predicted to begin further downstream compared to the results for $\lambda = \lambda_\text{opt}$ and the reference data.
For weaker regularization, here e.g. $\lambda = \num{1e-6}$, the $c_p$-distribution (dashed line) matches the reference data very closely with a decrease of cost function $\mathcal{I}_1$ down to $3.62\%$ of its initial value.
However, overfitting can be observed: In the pressure plateau, the reference data include slight fluctuations, to which the inversion result with the weak regularization tries to fit as well.
These fluctuations most probably stem from imprecisions during the process of injecting the reference data and should hence not be incorporated.
\begin{figure}[ht]
	\centering
	\def\cellwidth {.3\textwidth}
\def\mylinebreak {\vspace{2px}\\}
\begin{tabular}{ccc}
	$\lambda = \num{2e-3}$ &
	$\lambda = \lambda_\text{opt} = \num{5e-4}$ &
	$\lambda = \num{1e-6}$ 
	\mylinebreak
	\begin{minipage}{\cellwidth}
		\includegraphics[
			trim={150px 440px 990px 340px},
			clip,
			width=\textwidth
		]{%
			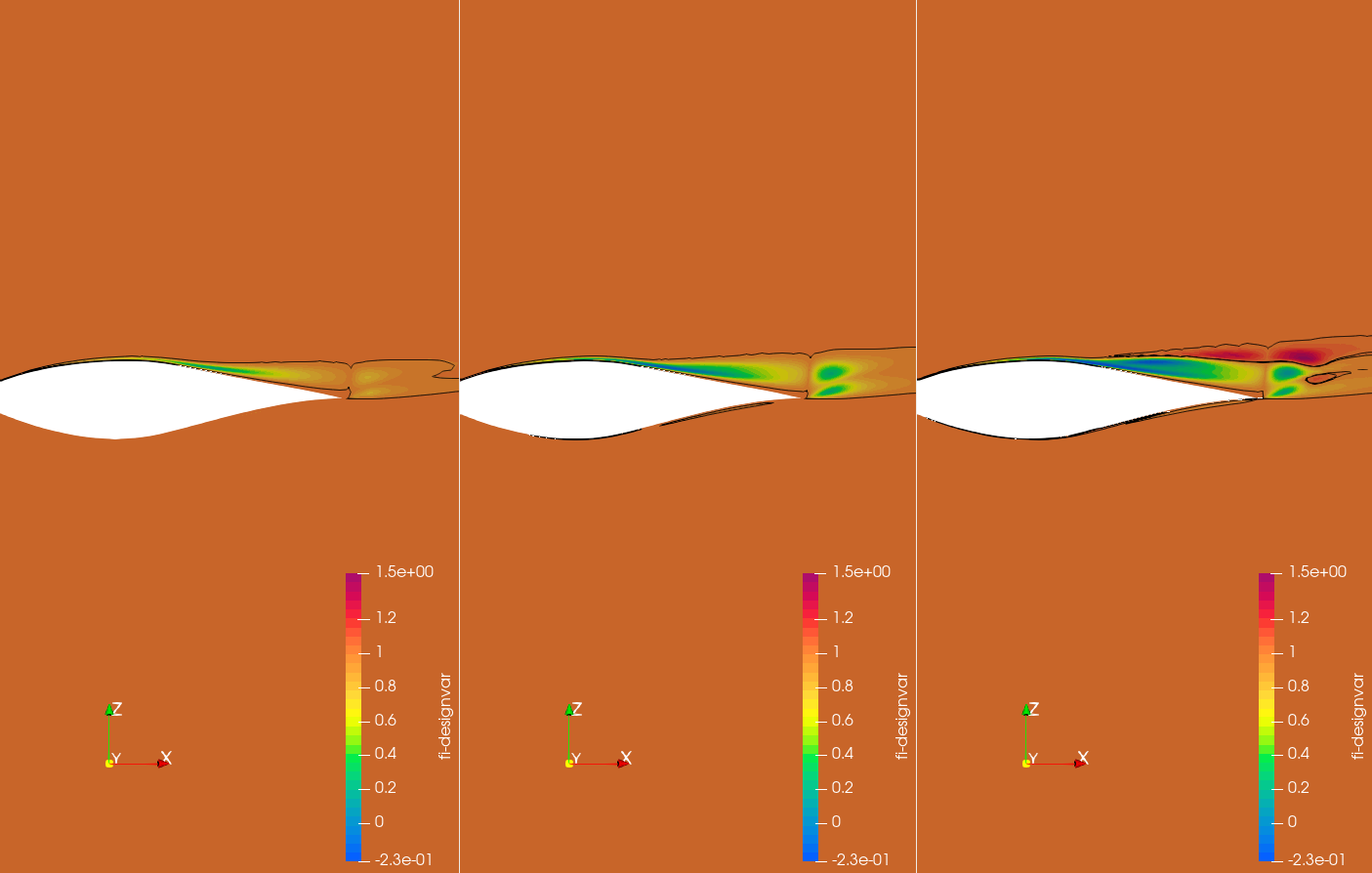%
		}
	\end{minipage}
	&		
	\begin{minipage}{\cellwidth}
		\includegraphics[
			trim={620px 440px 520px 340px},
			clip,
			width=\textwidth
		]{%
			figures/s809_reg_beta%
		}
	\end{minipage}
	&
	\begin{minipage}{\cellwidth}
		\includegraphics[
			trim={1090px 440px 50px 340px},
			clip,
			width=\textwidth
		]{%
			figures/s809_reg_beta%
		}
	\end{minipage}
\end{tabular}
\mylinebreak
\begin{minipage}{\textwidth}
	\centering
	\begin{tikzpicture}
		\pgfplotscolorbardrawstandalone[ 
			colormap name=pv-rainbow,
			colorbar sampled,
			samples=32,
			colorbar horizontal,
			point meta min=-0.23,
			point meta max=1.5,
			colorbar style={
				width=.6\textwidth,
				height=.7em,
				xtick={-0.23, 0.0, 0.5, 1.0, 1.5},
				xlabel=$\beta$,
				x label style={at={(axis description cs:-0.1,0.)},anchor=east}
			},
		]
	\end{tikzpicture}
\end{minipage}
	\caption{Resulting fields for $\beta$ for different values of the regularization parameter $\lambda$. The black contour marks $\beta = 1$.}
	\label{fig:s809_reg_beta}
\end{figure}

Resulting $\beta$-fields for all three cases are shown in Figure \ref{fig:s809_reg_beta}.
While the $\beta$-field for the stronger regularization ($\lambda = \num{2e-3}$) appears to be equivalent to a downscaled version of the $\beta$-field for $\lambda = \lambda_\text{opt}$, the $\beta$-field for the weak regularization ($\lambda = \num{1e-6}$) shows signs of overfitting by exhibiting more pronounced and additional extrema.
These findings confirm the relevance of choosing an optimum $\lambda$ and the L-Curve criterion as a method to do so.

\subsection{Grid Resolution}
In the next step, the influence of the grid resolution on the field inversion result is investigated.
Therefore, field inversions are conducted on six different grids at the same flow conditions as before.
The grids were obtained from \cite{s809-grids} and range from resolutions of $353 \times 49$ grid points to $2113 \times 289$ grid points, with $y^+_\text{max}$ ranging from $4.185$ to $0.799$.
The outer boundary of the computational domain has a distance of $1000c$ from the airfoil, where $c$ is the airfoil chord length.
Grid convergence with the baseline turbulence model is reached for the $1057 \times 145$ grid.

\begin{figure}%
	\def\cellwidth {0.32\textwidth}%
\def\mylinebreak {\vspace{2px}\\}%
\begin{minipage}{.9\textwidth}%
	\begin{tabular}{ccc}%
		$353 \times 49$ & $529 \times 73$ & $705 \times 97$
		\\%
		\begin{minipage}{\cellwidth}
			\includegraphics[
			trim={100px 750px 1020px 10px},
			clip,
			width=\textwidth
			]{%
				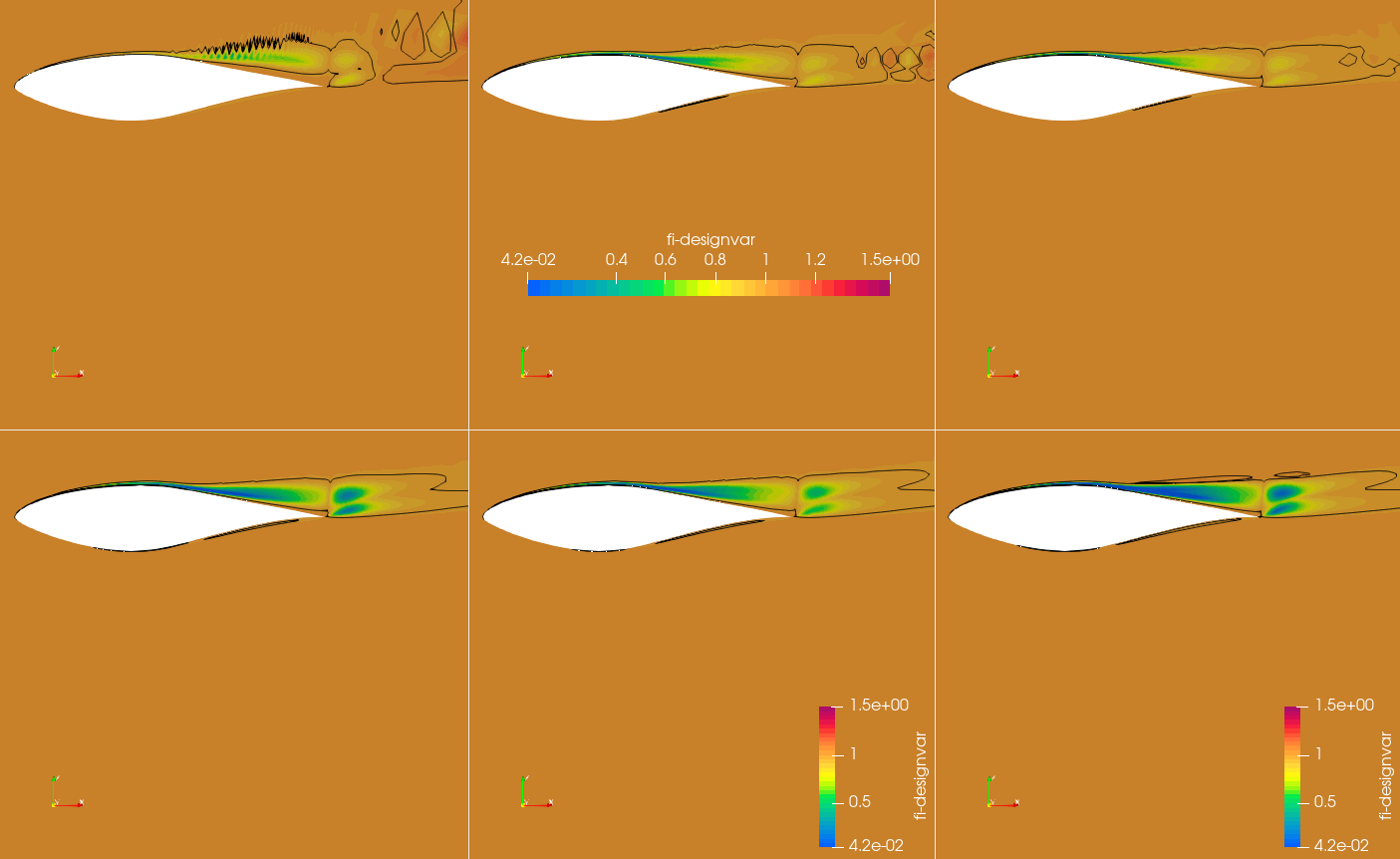%
			}
		\end{minipage}%
		&%
		\begin{minipage}{\cellwidth}
			\includegraphics[
			trim={570px 750px 550px 10px},
			clip,
			width=\textwidth
			]{%
				figures/s809_grid_beta%
			}
		\end{minipage}%
		&%
		\begin{minipage}{\cellwidth}
			\includegraphics[
			trim={1040px 750px 80px 10px},
			clip,
			width=\textwidth
			]{%
				figures/s809_grid_beta%
			}
		\end{minipage}%
		\mylinebreak%
		\begin{minipage}{\cellwidth}
			\includegraphics[
			trim={100px 320px 1020px 440px},
			clip,
			width=\textwidth
			]{%
				figures/s809_grid_beta%
			}	
		\end{minipage}%
		&%
		\begin{minipage}{\cellwidth}
			\includegraphics[
			trim={570px 320px 550px 440px},
			clip,
			width=\textwidth
			]{%
				figures/s809_grid_beta%
			}
		\end{minipage}%
		&%
		\begin{minipage}{\cellwidth}
			\includegraphics[
			trim={1040px 320px 80px 440px},
			clip,
			width=\textwidth
			]{%
				figures/s809_grid_beta%
			}	
		\end{minipage}%
		\mylinebreak
		$1057 \times 145$ & $1409 \times 193$ &	$2113 \times 289$
	\end{tabular}
\end{minipage}%
\begin{minipage}{.1\textwidth}
	\begin{tikzpicture}
	\pgfplotscolorbardrawstandalone[ 
		colormap name=pv-rainbow,
		colorbar sampled,
		samples=32,
		point meta min=0.,
		point meta max=1.5,
		colorbar style={
			width=.7em,
			height=8.5em,,
			ytick={0.0, 0.5, 1.0, 1.5},
			ylabel=$\beta$,
			y label style={at={(axis description cs:2.,0.5)},anchor=west, rotate=-90}
		},
	]
	\end{tikzpicture}
\end{minipage}%
	\caption{Field inversion results $\beta$ for different grids.}
	\label{fig:s809_grid_beta}
\end{figure}

For each grid, the optimal regularization parameter $\lambda_\text{opt}$ was determined separately as described in the previous section.
The corresponding results for $\beta$ are plotted in Figure \ref{fig:s809_grid_beta}.
On all grids, the field inversion process was successful and the cost function could be reduced adequately.
In general, the magnitude of the correction $\beta$ appears to increase with the grid resolution.
For the $1057 \times 145$ and the $1409 \times 193$ grid, almost no differences in magnitude for the $\beta$-field can be made out in the region of flow separation.
For the $2113 \times 289$ grid, additional maxima can be seen, qualitatively similar to those appearing due to over-fitting as found in the chapter before.

Since the field inversion shows good results on all grids, including under-resolved grids, it must be assumed that the obtained $\beta$-fields on the under-resolved grids also account for discretization errors.
Currently, we are interested only in improving the turbulence model itself and hence use only results from converged grids, but it can be reasoned that a FIML approach could be used to improve simulations on under-resolved grids in general as well.

\subsection{Area of Optimization}

\begin{figure}
	\begin{minipage}[t]{.54\textwidth}
	\raisebox{.07\textwidth}{
		\begin{tikzpicture}
			\pgfplotscolorbardrawstandalone[ 
			colormap name=pv-rainbow,
			colorbar sampled,
			samples=32,
			point meta min=0.,
			point meta max=1.5,
			colorbar style={
				width=.7em,
				height=9em,,
				ytick={0.0, 0.5, 1.0, 1.5},
				ylabel=$\beta$,
				y label style={at={(axis description cs:0.,1.1)},anchor=west, rotate=-90}
			},
			]
		\end{tikzpicture}%
	}%
	\raisebox{.12\textwidth}{
		\def\svgwidth{.8\textwidth}
		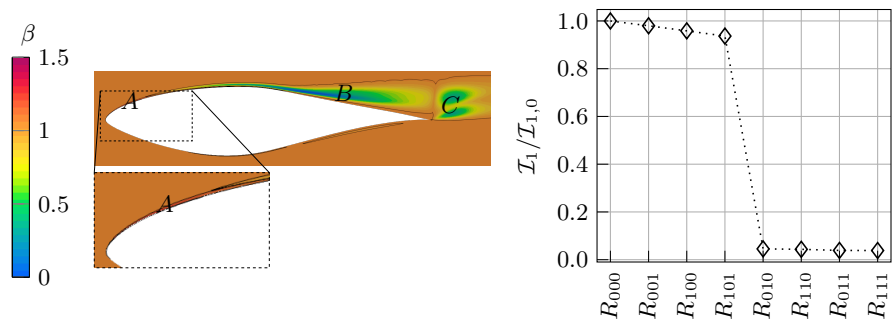
	}
\end{minipage}%
\hfill
\begin{minipage}[t]{.45\textwidth}
	\def\figwidth {\textwidth}
	\def\figheight {.9\textwidth}
\begin{tikzpicture}

\begin{axis}[
height=\figheight,
legend style={fill opacity=0.8, draw opacity=1, text opacity=1, draw=white!80!black},
tick pos=left,
width=\figwidth,
x grid style={white!69.0196078431373!black},
xmajorgrids,
xmin=-0.35, xmax=7.35,
xtick style={color=black},
xtick={0,1,2,3,4,5,6,7},
xtick={0,1,2,3,4,5,6,7},
xticklabel style = {rotate=90.0},
xticklabels={$R_{000}$,$R_{001}$,$R_{100}$,$R_{101}$,$R_{010}$,$R_{110}$,$R_{011}$,$R_{111}$},
xticklabels={\(\displaystyle R_{000}\),\(\displaystyle R_{001}\),\(\displaystyle R_{100}\),\(\displaystyle R_{101}\),\(\displaystyle R_{010}\),\(\displaystyle R_{110}\),\(\displaystyle R_{011}\),\(\displaystyle R_{111}\)},
y grid style={white!69.0196078431373!black},
ylabel={\(\displaystyle \mathcal{I}_{1}/\mathcal{I}_{1,0}\)},
ymajorgrids,
ymin=-0.00965702929441915, ymax=1.04788463310113,
ytick style={color=black},
ytick={-0.2,0,0.2,0.4,0.6,0.8,1,1.2},
yticklabels={−0.2,0.0,0.2,0.4,0.6,0.8,1.0,1.2}
]
\addplot [semithick, black, dotted, mark=diamond, mark size=3, mark options={solid,fill opacity=0}, forget plot]
table {%
0 0.999814557537697
1 0.97974937696262
2 0.958467224365461
3 0.936602640967776
4 0.0452619687277336
5 0.0434677116102359
6 0.0387970981762166
7 0.0384130462690149
};
\end{axis}

\end{tikzpicture}
\end{minipage}%
	\caption{
		Left: The three areas of interest marked by $A$, $B$ and $C$.
		Right: Results after activating $\beta$ in the different regions.
	}
	\label{fig:s809_regions}
\end{figure}

Finally, the influence of the correction term $\beta$ depending on the area where it is active is investigated.
As seen on the left in Figure \ref{fig:s809_regions}, three areas where $\beta$ has changed noticeably can be distinguished, marked with $A$, $B$ and $C$.
In region $A$, at the upper surface close to the leading edge, turbulent production was increased by increasing $\beta$.
In region $B$, spanning almost the entire upper surface except for the leading edge, turbulent production was reduced by decreasing $\beta$.
Region $C$, which encompasses the two spots corresponding to the vortices downstream of the trailing edge, again sees a reduction in turbulent production.

The influence of the correction term depending on the area where it is active is investigated by running RANS simulations with the augmented model starting with the full $\beta$-field as seen in Figure \ref{fig:s809_regions} and turning off the correction in the different areas one by one by setting $\beta$ to $1$.
The different tries are declared as $R_{ABC}$, where $A$, $B$ and $C$ are either $1$ or $0$, depending on whether the correction is active in the respective region or not.

The results are shown on the right in Figure \ref{fig:s809_regions}.
Turning the correction term off in all regions ($R_{000}$) corresponds to the baseline turbulence model, hence the cost function part $\mathcal{I}_1$ remains unchanged at $100\%$ of its initial value.
Activating the correction term in regions $A$ and $C$ decreases $\mathcal{I}_1$ only slightly to $98.0\%\,(R_{100}), 95.8\%\,(R_{001})$ and $93.7\%\,(R_{101})$ of its initial value.
The correction term in region $B$ has by far the largest influence by reducing $\mathcal{I}_1$ to $4.53\%\,(R_{010})$ of the initial value alone and to $4.35\%\,(R_{110}), 3.88\%\,(R_{011})$ and $3.84\%\,(R_{111})$ when active in conjunction with regions $A$ and $C$.
This behaviour was expected however since region $B$ covers the largest area and especially the area where flow separation occurs.%
\section{Application}
\label{chap:application}

\subsection{Training}

In this section the entire FIML approach is applied.
First, training data is generated by conducting field inversions on the S809 airfoil for angles of attack $\alpha = 6.2^\circ, 10.2^\circ$ and $20.1^\circ$.
As discussed in the previous chapter, for each case the regularization parameter $\lambda$ is determined separately and the field inversions are conducted only on the fully resolved $1057 \times 145$ grid to ensure optimal training data.
From the complete set of data, the samples in the freestream, where the turbulence model is not active, are discarded together with $99.5\%$ of the remaining samples where $\beta = 1$.
This is done to reduce the frequency at which $\beta = 1$ appears in the training data as otherwise the machine learning model would simply learn to always predict $\beta = 1$.
The remaining samples are split into a training set totaling approximately $38000$ samples and a testing set totaling approximately $4 000$ samples.

Second, a machine learning model is trained to model $\beta$ according to Eq. (\ref{eq:ml-beta}).
The neural network used in this case is a fully connected network with two layers with 200 neurons each.
Dropout layers are applied after each hidden layer with a drop-out rate of $0.2$.
The mean squared error is used as loss function and the network is trained for $500$ epochs using the Adam optimizer.
The dimensionless flow features $\eta_i(\mathbf{U}, \tilde{\nu})$ used as input are
\begin{equation}
	\eta_1 = \log (\frac{P}{D}), \eta_2 = \log(\chi), \eta_3 = \log(\frac{|S|}{|\Omega|}) \,\text{and}\, \eta_4 = \log(\frac{\mu_t |S|}{\tau_w}),
\end{equation}
similar to the features presented in \cite{holland}.
The features represent the ratio of the production and the destruction term of the SA model, the non-dimensionalized SA transport variable, the ratio of the strain and vorticity magnitudes and the ratio of the local turbulent shear stress to the shear stress at the wall respectively.
The logarithms are applied due to the fact that each features' values span multiple orders of magnitudes.

\subsection{Numerical Results}

Finally, RANS simulations with the ML-augmented turbulence model are conducted for multiple angles of attack from $\alpha = 0.0^\circ$ up to $\alpha = 20.1^\circ$.
\begin{figure}
	\centering
	\begin{minipage}[t]{.5\textwidth}
		\def\figwidth {\textwidth}
		\def\figheight {\textwidth}
\begin{tikzpicture}

\begin{axis}[
height=\figheight,
legend cell align={left},
legend cell align={right},
legend plot pos=right,
legend style={fill opacity=0.8, draw opacity=1, text opacity=1, at={(0.97,0.03)}, anchor=south east, draw=white!80!black},
tick pos=left,
width=\figwidth,
x grid style={white!69.0196078431373!black},
xlabel={\(\displaystyle \alpha \, [^\circ]\)},
xmajorgrids,
xmin=-2.29871903821714, xmax=21.4827517133723,
xtick style={color=black},
y grid style={white!69.0196078431373!black},
ylabel={\(\displaystyle c_l\, [-]\)},
ymajorgrids,
ymin=-0.0934668234515587, ymax=1.57296447147502,
ytick style={color=black}
]
\addplot [only marks, mark=x, draw=black, fill=black, colormap/viridis]
table{%
x                      y
0 0.133266370230574
2.1 0.374036936648379
4.2 0.610638079852547
6.2 0.829036674441379
8.2 1.03690000157752
10.2 1.22843522169155
12.2 1.38932289318005
14.2 1.45756242401616
16.2 1.44366386405543
18.2 1.4707626634863
20.1 1.47882171510987
};
\addlegendentry{SA-neg}
\addplot [only marks, mark=o, draw=black, colormap/viridis]
table{%
x                      y
6.16 0.830812230400002
10.2 1.0312254614679
20.1 0.953775617426945
};
\addlegendentry{Inversion}
\addplot [only marks, mark=triangle*, draw=black, fill=black, colormap/viridis]
table{%
x                      y
0 0.13331109768136
2.1 0.375469343414044
4.2 0.613299246397271
6.16 0.828754229735657
8.2 1.03949990474843
10.2 1.13907541463107
12.2 1.06491457489465
14.2 1.11865539845461
16.2 1.13392694957396
18.2 1.06601601074131
20.1 0.963745866347611
};
\addlegendentry{Augmented SA-neg}
\addplot [only marks, mark=diamond, draw=black, colormap/viridis]
table{%
x                      y
-0.958007871453058 0.0221118320266744
0.0354792819890957 0.144611156168334
1.05746895644848 0.264294403892944
2.07925357679982 0.385385689826079
3.04485337154886 0.506476975759214
4.17859753488872 0.630384338109399
4.60622363733252 0.691132060633672
4.87820512785461 0.721281026904169
5.08575121289961 0.747355808543517
5.40745357799794 0.770986079404177
5.58355328652097 0.789727328707459
6.16382793253014 0.823950479609104
7.16144939449314 0.894841292191083
8.20724907859924 0.96980628940421
9.19205465880989 1.01706683112553
10.2338197134754 1.008103624937
11.1813643041468 0.976324984814047
11.1813643041468 0.976324984814047
12.1842070512682 1.01136297264192
13.2181401781962 1.05617900358455
14.2056750665578 1.08469829600259
15.2599000062604 1.10180987145341
16.2074445969318 1.07003123133046
17.2417337205751 1.00077009260094
18.2073154781573 0.956768898584536
19.1902224092194 0.905434172232068
20.1364616786012 0.882618738297638
};
\addlegendentry{Somers \cite{s809}}
\end{axis}

\end{tikzpicture}%
	\end{minipage}%
	\hfill
	\begin{minipage}[t]{.5\textwidth}
		\def\figwidth {\textwidth}
		\def\figheight {\textwidth}
		\input{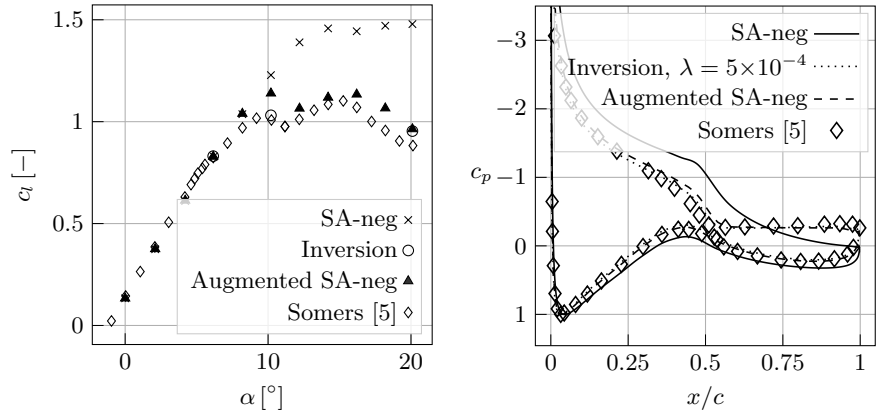}%
	\end{minipage}%
	\caption{Left: Lift coefficients for the S809 airfoil at Re = $2\times 10^6$. 
	Right: $c_p$-distributions for $\alpha = 12.2^\circ$ }
	\label{fig:results_ml}
\end{figure}%

Figure \ref{fig:results_ml} compares results from the augmented RANS model, the original turbulence model, the field inversion and the reference data.
For $\alpha \lessapprox 6.2^\circ$, the agreement between the measured lift and the predicted lift using the original RANS model is good, while it quickly worsens for larger $\alpha$ as soon as flow separation occurs.
The lift predictions of the field inversion results agree much better, matching the reference lift with deviations of only $0.8\%$ ($\alpha = 6.2^\circ$), $2.3\%$ ($\alpha =10.2^\circ$) and $8\%$ ($\alpha = 20.1^\circ$).
The predictions of the ML-augmented turbulence model remain unchanged in the linear area of the lift curve, which is intended since the results of the original turbulence model are already satisfying here.
For higher $\alpha$, the augmentation improves the turbulence model noticeably as the lift is predicted far more accurately.
The maximum deviation of the augmented model is considerably lower with $13\%$ (at $\alpha = 10.2^\circ$) as opposed to a maximum deviation of $> 50\%$ (at $\alpha = 20.1^\circ$) of the baseline model.
\begin{figure}
	\def\cellwidth {0.49\textwidth}
\def\mylinebreak {\vspace{2px}\\}
\begin{minipage}{.1\textwidth}
	\begin{tikzpicture}
	\pgfplotscolorbardrawstandalone[ 
	colormap name=pv-rainbow,
	colorbar sampled,
	samples=32,
	point meta min=-0.16,
	point meta max=1.7,
	colorbar style={
		width=1em,
		height=2\textwidth,,
		ytick={-0.16, 1.0, 1.7},
		ylabel=$\beta$,
		y label style={at={(axis description cs:0.,1.2)},anchor=west, rotate=-90}
	},
	]
	\end{tikzpicture}
	\vspace{1em}
\end{minipage}%
\hfill
\begin{minipage}{.75\textwidth}
	\begin{tabular}{cc}
		Inversion, $\beta$
		&
		\parbox{\cellwidth}{\centering
			Baseline,\\ Ma + stream lines
		}
		\\
		\begin{minipage}{\cellwidth}
			\includegraphics[
			trim={50px 550px 750px 150px},
			clip,
			width=\textwidth
			]{%
				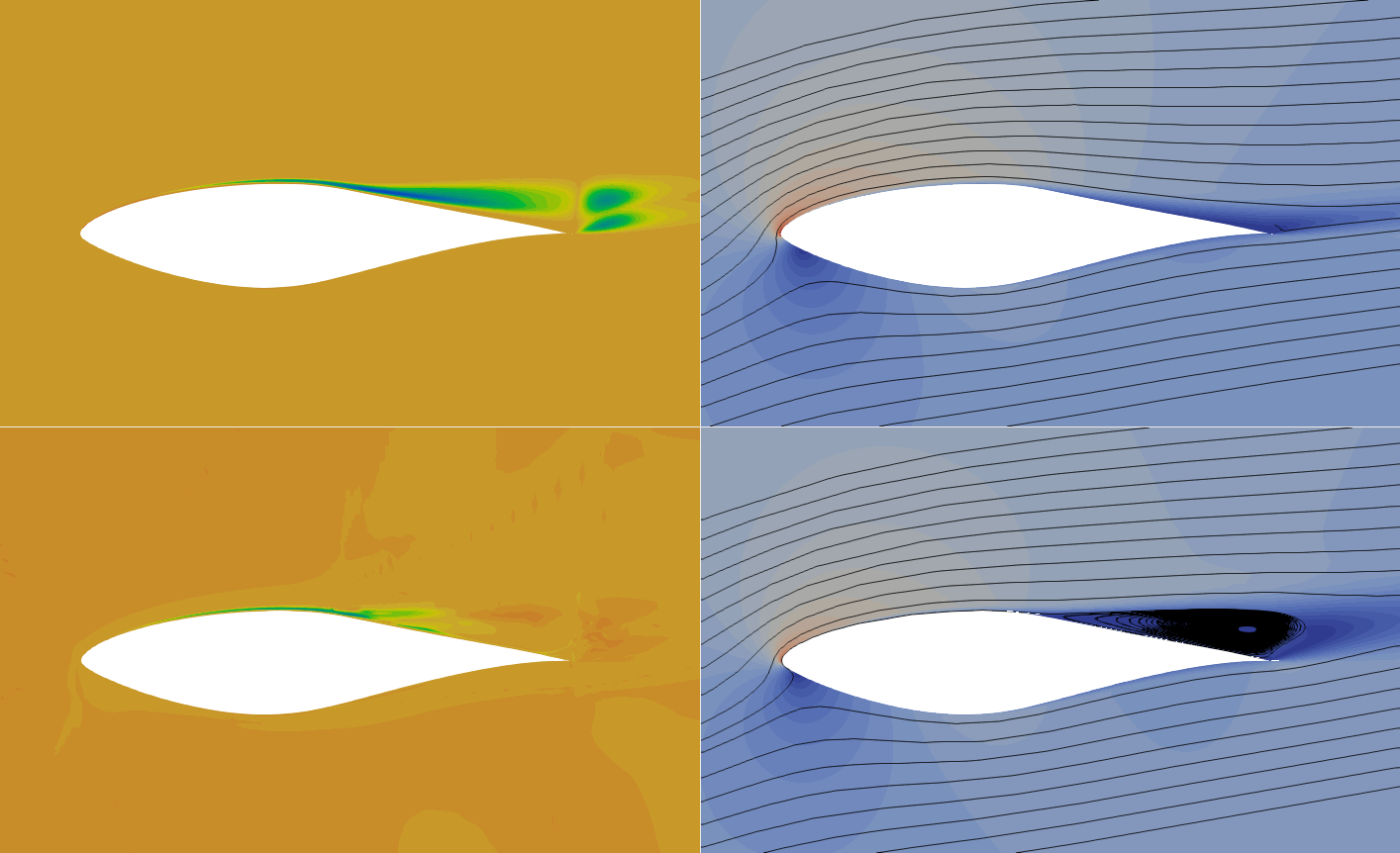%
			}
		\end{minipage}%
		&%
		\begin{minipage}{\cellwidth}
			\includegraphics[
			trim={760px 550px 40px 150px},
			clip,
			width=\textwidth
			]{%
				figures/s809_fiml%
			}
		\end{minipage}%
		\mylinebreak%
		\begin{minipage}{\cellwidth}
			\includegraphics[
			trim={50px 120px 750px 580px},
			clip,
			width=\textwidth
			]{%
				figures/s809_fiml%
			}
		\end{minipage}%
		&%
		\begin{minipage}{\cellwidth}
			\includegraphics[
			trim={760px 120px 40px 580px},
			clip,
			width=\textwidth
			]{%
				figures/s809_fiml%
			}	
		\end{minipage}%
		\mylinebreak
		ML-augmented, $\beta$
		&
		\parbox{\cellwidth}{\centering
			ML-augmented,\\ Ma + stream lines
		}
	\end{tabular}
\end{minipage}%
\hfill
\begin{minipage}{.1\textwidth}
	\begin{tikzpicture}
	\pgfplotscolorbardrawstandalone[ 
	colormap name=pv-bluewhitered,
	colorbar sampled,
	samples=32,
	point meta min=0.,
	point meta max=0.3,
	colorbar style={
		width=1em,
		height=2\textwidth,,
		ytick={0.0, 0.1, 0.2, 0.3},
		ylabel=Ma,
		y label style={at={(axis description cs:0.,1.2)},anchor=west, rotate=-90}
	},
	]
	\end{tikzpicture}
	\vspace{2em}
\end{minipage}%
	\caption{%
		Left: $\beta$-field from field inversion and as predicted by the ML-model.
		Right: Mach number and stream lines from the baseline turbulence model and the ML-augmented model.
	}
	\label{fig:result_fiml}
\end{figure}

On the right in Figure \ref{fig:results_ml}, the improvement of the predictions due to the ML-augmentation is illustrated based on the $c_p$-distribution exemplary for $\alpha = 12.2^\circ$, for which the field inversion was done but not included in the training data for the ML model.
While the $c_p$-distribution of the augmented model (dashed line) doesn't come as close to the experimental values as the field inversion result (dotted), it still represents a large improvement with respect to the baseline model (solid).
Figure \ref{fig:result_fiml} compares the results for the $\beta$-field as computed during field inversion and as predicted by the ML-augmentation.
While the inversion result is smoother in general and shows stronger modifications, the presumably most import area leading up to the point of flow separation shortly after the thickest airfoil section, is predicted similarly by the ML-model.
In the right half of Figure \ref{fig:result_fiml}, the Ma-number and the streamlines of the solution are plotted for the original turbulence model and the ML-augmented model.
While the original model shows only a very small region of flow separation, the ML-augmented model develops a much larger separation bubble as expected.%
\section{Conclusion}
\label{chap:conclusion}
The field inversion and machine learning approach was successfully reproduced based on the DLR TAU-code and the negative Spalart-Allmaras model.
The influence of the regularization parameter, the grid resolution and the regions where $\beta$ is optimized during field inversion on the field inversion results were investigated based on the flow around the S809 airfoil with flow separation at the trailing edge.
For the regularization, a method to determine the optimal regularization parameter was demonstrated and the influence a suboptimal choice can have was discussed.
It was shown and discussed that the field inversion can also compensate for spatial discretizaton, returning correct results on coarse grids as well.
Different regions where the optimization is active were detected and their influence on the end result was investigated.
Under consideration of these information, a machine learning model was trained using data from field inversions of the S809 at multiple angles of attack.
The RANS turbulence model was then augmented by the machine learning model and improved results were shown on the S809 airfoil.

A future step in the context of the FIML approach is to include inversion results from additional test cases with different geometries and at different flow conditions in the training data, which is expected to be essential to be able to provide a robust and reliable augmentation.%
%
%
%

%
%
\end{document}